\documentstyle[12pt]{article} 
\font\tenrm=cmr10 

\def\lrb#1{ \left( { #1 } \right) } 

\def\gam{\gamma} 
\def\gab{\bar \gamma} 
\def\ya{y_1} 
\def\yb{y_2} 
\def\yc{y_3} 
\def\za{z_1} 
\def\zb{z_2} 
\def\zc{z_3} 
\def\nubarr{ { {n_\nu} \over {n_{\bar \nu}} } } 
\def\nubarrs{ {{n_\nu}   /   {n_{\bar \nu}} } } 
\def\ka{k_1} 
\def\kab{k_{12}} 
\def\dyads{ {{dy_1} \over {d \tau }} }

\def\dya2ds2{ {{d^2 y_1} \over {d \tau^2 }} }
\def\yasol{ y_1^{sol} }
\def\cs{ \cos{ 2 \theta} } 
 
\def\Acoeffh{ \hat A} 
\def\Bcoeffh{ \hat B} 
\def\Ccoeffh{ \hat C} 
\def\Acoeff{ A } 
\def\Bcoeff{ B } 
\def\Ccoeff{ C } 
\def\var{ y_1 }
\def\half{ { {1} \over {2} } } 
\def\xp{ y_{+} } 
\def\xm{ y_{-} } 
\def\xo{ y_{0} } 
\def\qs{ q^2 } 
\def\s{\tau} 
\def\sqr{ \sqrt{ { \Bcoeff}^2 - 4  \Acoeff \Ccoeff}  } 
\def\JacobiSN{ sn } 
\def\JacobiDN{ dn } 
\def\EllipticK{ K } 
\def\dphids{ {{d \varphi } \over {d \tau }} } 
\def\dphidsbar{ {{d \bar \varphi } \over {d \tau }} } 
\def\factora{ \half \lrb{{1 - {{{ \ka}}^2} + {{\lrb{ \nubarr}}^2}}}} 
\def\factorb{ {{ \ka} \over {2 { {\gam}^2} }} {{ \left( 1 + {{{ \ka}}^2} - 
    {{\lrb{ \nubarr}}^2} \right) } }} 
\def\factorc{ - \half \lrb{{1 + {{{ \ka}}^2} - {{\lrb{ \nubarr}}^2}}}} 

\def\kap{\kappa} 
\def\De{\Delta} 
\def\th{\theta} 
\def\fr#1#2{{{#1} \over {#2}}} 
\def\expect#1{\langle{#1}\rangle} 
\def\eq{Eq.\ } 
\def\eqs{Eqs.\ } 
\def\ie{i.e.} 
\def\ctth{\cos{ 2\theta }} 
\def\stth{\sin{ 2\theta }} 
\def\sttth{\sin^2{ 2\theta }} 
 
\def\ddtau#1{{{d #1} \over {d \tau}}} 
\def\secttit#1{\vglue 0.6cm{\bf\large\noindent{#1}}\vglue 0.4cm} 

\def\kapo{\fr{\Delta}{2 \sqrt{2} G_F E_0 n_\nu}} 
\def\taufac{\half\fr{\Delta}{E_0}\sqrt{\fr{1}{\kappa_0}}} 
\def\invtaufac{2\fr{E_0}{\Delta}\sqrt{\kappa_0}}

\def\theequation{\thesection.\arabic{equation}}
\def\@eqnnum{{\rm (\theequation)}} 
\def\secttit#1{\vglue 0.6cm{\bf\large\noindent{#1}}\vglue 0.2cm} 

\newcommand{\bref}[1]{(\ref{#1})} 
\newcommand{\ct}[1]{\cite{#1}}

\newcommand{\beq}{\begin{equation}} 
\newcommand{\eeq}{\end{equation}} 
\newcommand{\bea}{\begin{eqnarray}}   
\newcommand{\eea}{\end{eqnarray}}  
  
\renewenvironment{thebibliography}[1]  
 { \rm  
   \begin{list}{\arabic{enumi}.}  
    {\usecounter{enumi} \setlength{\parsep}{0pt}  
     \setlength{\itemsep}{3pt} \settowidth{\labelwidth}{#1.}  
     \sloppy  
    }}{\end{list}}

\begin{document}  
\titlepage  
  
\begin{flushright}  
{MPI-PhT/95-57\\}  
{CCNY-HEP-95/5\\}   
{hep-ph/9604341\\}  
{July 1995\\}  
\end{flushright}  
\vglue 1cm

\begin{center}   
{ 
{\Large \bf Bimodal Coherence in \\  
Dense Self-Interacting Neutrino Gases   
\\}  
\vglue 1.0cm  
{Stuart Samuel$^{*}$\\}   
\bigskip  
{\it Max-Planck-Institut f\"ur Physik\\}
{\it Werner-Heisenberg-Institut\\} 
{\it F\"ohringer Ring 6\\} 
{\it 80805 Munich, Germany\\} 
  
\vglue 0.8cm  
} 
 
\vglue 0.3cm  
  
{\bf Abstract} 
\end{center}  
{\rightskip=3pc\leftskip=3pc\noindent  
Analytical solutions are obtained 
to the nonlinear equations describing 
neutrino oscillations 
when explicit neutrino-antineutrino 
asymmetries are present.  
Such a system 
occurs in the early Universe 
if neutrinos have a non-zero chemical potential. 
Solutions to the equations lead to 
a new type of coherent behavior 
governed by two modes.  
These bimodal solutions provide new insights 
into dense neutrino gases 
and into neutrino oscillations in the early Universe, 
thereby allowing one to surmise the flavor behavior 
of neutrinos with  
a non-zero chemical potential.  

}

\vfill
 
\textwidth 6.5truein
\hrule width 5.cm
\vskip 0.3truecm 
{\tenrm{
\noindent 
$^*$Permanent address and address after August, 1995:\\   
\hspace*{0.2cm}Physics Department, City College of New York, 
New York, NY 10031, USA.\\
\hspace*{0.2cm}E-mail: samuel@scisun.sci.ccny.cuny.edu\\}}
 
\newpage  
  
\baselineskip=20pt  
 
{\bf\large\noindent I.\ Introduction}  
\vglue 0.4cm   
\setcounter{section}{1}   
\setcounter{equation}{0}   

In this work we undertake 
a theoretical study of flavor oscillations 
in dense neutrino gases.  
Such gases 
appear in physically interesting systems.  
An example occurs  
during the early Universe 
when neutrinos are self-interacting 
and fill space densely 
\ct{kt,kim}.    
Likewise, during the final collapse 
of a supernova, 
neutrinos are emitted copiously 
\ct{kim,bahcall}.  
If neutrinos have masses and mix 
then oscillations among neutrino flavors 
can effect the physics of these systems.  

The treatment of neutrino oscillations in a dense gas 
is not so straightforward.  
The flavor behavior of a particular neutrino 
depends on the flavor content of background neutrinos.  
However, the background neutrinos also oscillate.  
To know the flavor content of background neutrinos, 
it is necessary to know the oscillations 
of all individual neutrinos.  
Nonetheless, 
a Hartree-Fock-like self-consistent formalism 
has appeared which can handle the system 
\ct{samuel93a}.  
The behavior of the gas 
depends on certain statistical properties 
such as the energy distribution, 
the nature of density perturbations 
and the initial production of neutrinos.  
For the case in which only an energy distribution is involved,  
the formalism has been further developed 
for the neutrino-antineutrino gas 
in refs.\ \ct{sr93a,kps93a,ks94a}.  
The physical effects of neutrino oscillations 
on supernovae is treated  
in ref.\ \ct{qf95a}.  

Since the system is self-interacting 
and nonlinear, 
one usually has to resort to numerical methods 
to determined the physics and flavor behavior.  
This approach was used 
in refs.\ \ct{kps93a,ks94a,ks93a,ks95a,ks95b} 
to analyze neutrino oscillations 
in the early Universe 
for the case 
in which neutrinos have a chemical potential $\mu_\nu$  
which is zero.  
In a gas for which $\mu_\nu = 0$,  
the total number of antineutrinos 
is equal to the total number of neutrinos.  
It is unknown whether $\mu_\nu = 0$ 
in the early Universe.  
Indeed, the excess of electrons over positrons 
implies that the chemical potential 
for charged leptons is nonzero.  
Because this excess is tiny, 
being related to the baryon asymmetry of the Universe, 
the chemical potential for electrons is quite small.  
Nevertheless, 
because neutrinos are so dense 
at one second after the Big Bang, 
when the temperature of the Universe is about an MeV, 
even a tiny $\mu_\nu$ 
can have an effect. 
One purpose 
of the current work 
is to surmise 
the flavor behavior of neutrino oscillations 
in the early Universe 
when $\mu_\nu \ne 0$.  
Various aspects of neutrino oscillations 
in the early Universe are also addressed in 
refs.\ [12--20].

Thus one is led to consider 
the system when the total  number of antineutrinos 
is not necessarily equal to 
the total number of neutrinos.  
We assume the gas is homogeneous and isotropic.  
Under this assumption,  
the averaged spatial neutrino currents 
are zero and one needs only to consider neutrino densities.  
We also assume that the energy distribution 
is the most important statistical property 
of the gas.  
For simplicity, 
two flavors, electron and muon, are treated.  
In many regions of parameter space, 
the three-flavor case is accurately approximated 
by the two-flavor case. 
Finally, 
we consider situations in which 
hard scattering processes 
are much smaller than forward-scattering phase effects 
so that hard scattering can be and is ignored.  
This occurs for times greater than  
one second after the Big Bang 
and outside the neutrino sphere of a supernova. 
In these situations, 
$G_F E^2 \ll 1 $,  
where 
$ 
  G_F \simeq 1.17 \times 
10^{-11} \, {\rm MeV}^{-2}  
$ 
is the Fermi coupling constant 
and $E$ is the energy of a typical neutrino. 
The analytic results as well as the graphs displayed 
in the figures hold only for the case 
in which non-forward scattering can be neglected.

Several parameters describe the system.  
Two important ones are 
$\Delta$, which is the mass squared difference 
between the two mass-eigenstate neutrinos, 
and $\theta$, 
which is the vacuum mixing angle: 
\beq
  \Delta = \lrb{m_2^2 - m_1^2} 
\quad , 
\label{1p1}  
\eeq 
and 
\beq
 \nu_1 = 
  \nu_{e L} \cos{\th} - \nu_{\mu L} \sin{\th}  
\quad , \quad \quad 
 \nu_2 = 
  \nu_{e L} \sin{\th} + \nu_{\mu L} \cos{\th}   
\quad . 
\label{1p2}  
\eeq 
Here, 
$\nu_1$ and $\nu_2$ are the mass-eigenstate neutrinos, 
which have masses $m_1$ and $m_2$,  
and $\nu_{e L}$ and $\nu_{\mu L}$ 
are the left-handed electron and muon neutrinos, 
which are the states entering in the currents of 
the weak interactions. 

Let $N_\nu$ and $N_{\bar \nu}$ be respectively 
the number of neutrinos and the number of antineutrinos 
in the gas.  
We use labels $j$ and $k$ to enumerate 
respectively the neutrinos and the antineutrinos.  
Thus the index $j$ runs from $1$ to $N_\nu$ 
and the index $k$  
runs from $1$ to $N_{\bar \nu}$. 
The energy of the $j$th neutrino 
and the $k$th antineutrino 
is denoted by $E^j$ and $\overline E^k$.    
The system is put in a box of volume ${\cal V}$.  
The neutrino and antineutrino densities 
$n_{\nu}$ and $n_{\bar \nu}$ 
are given by   
$n_{\nu} =  N_{\nu} / {\cal V}$ and 
$n_{\bar \nu} =  N_{\bar \nu} / {\cal V}$.  

The wave function for $j$th neutrino is governed by a
two-component vector $\nu^j$ in flavor space, \ie, 
\beq 
  \nu^j  = 
\lrb{\matrix{ 
     \nu_e^j   \cr  
     \nu_\mu^j \cr}}  
\quad , 
\label{1p3} 
\eeq 
where $\nu_e^{j*} \nu_e^j$ 
(respectively, $\nu_\mu^{j*} \nu_\mu^j$) 
is the probability 
that the $j$th neutrino is an electron neutrino 
(respectively, muon neutrino).  
Likewise, for the $k$th antineutrino, 
the flavor wave function is
\beq 
  \bar \nu^k  = 
\lrb{\matrix{ 
     \bar \nu_e^k   \cr  
     \bar \nu_\mu^k \cr}}  
\quad .  
\label{1p4} 
\eeq 
Since, under the assumption of no hard scattering, 
neutrinos and antineutrinos 
are not created or destroyed but merely 
change flavor, 
conservation of probability implies that 
\beq 
  \nu_e^{j*} \nu_e^j + \nu_\mu^{j*} \nu_\mu^j = 1 
\ , \quad \quad 
  \bar \nu_e^{k*} \bar \nu_e^k + 
  \bar \nu_\mu^{k*} \bar \nu_\mu^k = 1 
\quad . 
\label{1p5}
\eeq

The equations governing neutrino oscillations 
are easier to understand intuitively and visually 
by using vectors $\vec v^j$ and $\vec w^k$ 
associated with \eqs\bref{1p3} and \bref{1p4} 
given by   
\ct{stodolsky87a}  
$$  
 \vec v^j \equiv 
          \lrb{\nu_e^{j*}\nu_e^j - \nu_\mu^{j*}\nu_\mu^j, 
 2 {\rm Re} ( {\nu_e^{j*}\nu_\mu^j}), 
 2 {\rm Im} ( {\nu_e^{j*}\nu_\mu^j}) } 
\quad ,  
$$ 
\beq 
 \vec w^k \equiv 
          \lrb{\bar \nu_e^{k*} \bar \nu_e^k - 
               \bar \nu_\mu^{k*} \bar \nu_\mu^k, 
 2 {\rm Re} ( {\bar \nu_e^{k*} \bar \nu_\mu^k}), 
 2 {\rm Im} ( {\bar \nu_e^{k*} \bar \nu_\mu^k}) } 
\quad ,  
\label{1p6} 
\eeq 
where ${\rm Re}$ and ${\rm Im}$ 
denote the real and imaginary parts 
of a complex number.  

The equations governing 
the flavor behavior of a gas 
of self-interacting neutrinos and antineutrinos 
are 
\ct{samuel93a,sr93a,kps93a,mt94a}
\beq  
 {{d\vec v^j} \over {dt}} = \vec v^j \times \vec B_v^j
\ , \quad \quad  
 {{d\vec w^k} \over {dt}} = \vec w^k \times \vec B_w^k
\quad ,  
\label{1p7} 
\eeq 
where 
$\vec B^j_v$ and $\vec B^k_w$
are given by 
\beq
  \vec B_v^j = 
   {{\vec \De } \over {2E^j}} - 
    {\vec V_{\nu \nu}} 
\ , \quad \quad 
 \vec B_w^k =  
   {{\vec \De } \over {2\overline E^k}} + 
    {\vec V^{*}_{\nu \nu}}  
\quad ,    
\label{1p8} 
\eeq 
with  
\beq 
  \vec \De \equiv  
     \De \lrb{ \ctth, -\stth, 0 }  
\quad .   
\label{1p9} 
\eeq
Here,    
\beq
    \vec V_{\nu \nu}   = 
   {\fr{\sqrt 2 G_F}{\cal V}} 
     \lrb{ \expect{\vec v} - \expect{\vec w^*}} 
\quad    
\label{1p10}
\eeq 
implements the interactions 
of neutrinos and antineutrinos 
among themselves.  
An asterisk on a vector indicates 
a change in sign of the third component 
so that  
$\vec w^* = \lrb{w_1, w_2, -w_3}$.  
In \eq\bref{1p10}, 
\beq 
  \expect{\vec v} = \sum\limits_j {\vec v^j} 
\ , \quad \quad 
  \expect{\vec w} = \sum\limits_k {\vec w^k}
\quad . 
\label{1p11} 
\eeq   
The neutrino-neutrino term $\vec V_{\nu \nu}$ 
makes the system of equations 
in \eq\bref{1p7} 
nonlinear.   

The equations in \eq\bref{1p7} 
resemble the motion of charged particles 
in ``magnetic fields'' $\vec B_v^j$ and $\vec B_w^k$.

The conservation of individual neutrinos 
in \eq\bref{1p5} 
corresponds to  
\beq 
 \vec v^j \cdot \vec v^j (t) = 1 
\ , \quad \quad 
 \vec w^k \cdot \vec w^k (t) = 1 
\quad . 
\label{1p12} 
\eeq 
The total number of neutrinos $ N_{\nu} $ 
and antineutrinos $ N_{\bar \nu}$  
is thus $ N_{\nu} = \sum_j | \vec v^j | $ 
and $N_{\bar \nu} = \sum_k | \vec w^k | $. 

If, at time $t=0$,
the $j$th neutrino and $k$th antineutrino 
begin as an electron neutrino and an electron antineutrino  
then 
\beq  
 \vec v^j ( 0 ) = \lrb{ 1, 0, 0 } 
\ , \quad \quad   
 \vec w^k ( 0 ) = \lrb{ 1, 0, 0 } 
\quad .  
\label{1p13} 
\eeq 
If they begin as a muon neutrino and muon antineutrino 
then 
$ \vec v^j ( 0 ) = \lrb{ -1, 0, 0 } $ 
and 
$ \vec w^k ( 0 ) = \lrb{ -1, 0, 0 } $.  
If there are both electron and muon neutrinos initially, 
then only the excess is relevant for neutrino oscillations.  
Thus if the there are more electron neutrinos than 
muon neutrinos at $t=0$ then 
\eq\bref{1p13} is general.  
In the current work, 
the initial conditions 
in \eq\bref{1p13} 
are referred to as {\it initial flavor eigenstates}. 

\secttit{II.\ Reduction of the Problem}  
\setcounter{section}{2}   
\setcounter{equation}{0}   

The ratio of the strength of the vacuum term  
to the maximum possible neutrino-interaction term 
for the $j$th neutrino and $k$th antineutrino 
is given by the parameters 
$\kap^j$ and $\bar \kap^k$, 
where 
\beq  
      \kap^j = 
  \fr{\De}{ \sqrt{2} G_F E^j \lrb{n_{\nu} + n_{\bar \nu} } } 
\ , \quad \quad 
 \bar \kap^k = 
  \fr{\De}{ \sqrt{2} G_F \overline E^k \lrb{n_{\nu} + n_{\bar \nu} } } 
\quad . 
\label{2p1} 
\eeq  
When $\kap^j \ll 1$ and $\bar \kap^k \ll 1$,  
the vacuum term is much smaller than the 
nonlinear neutrino term.  
This corresponds to the case 
when the neutrino gas is dense.  
Under these circumstances, 
ref.\ \ct{ks94a} 
found that alignment holds.  
Alignment has been observed numerically 
\ct{ks94a} 
and is understood theoretically 
\ct{ks94b}.  
When alignment holds, 
\beq  
  \vec v^j (t) \approx 
  \fr{\expect{\vec v}}{N_\nu} \equiv 
   \vec r_{v} (t) 
\ , \quad \quad 
  \vec w^k (t) \approx 
  \fr{\expect{\vec w}}{N_{\bar \nu}} \equiv 
   \vec r_{w} (t) 
\quad 
\label{2p2} 
\eeq  
are good approximations.  
Throughout the rest of this work, 
the dense neutrino case is treated,  
so that 
\eq\bref{2p2} hold quite accurately.   

When \eq\bref{2p2} is an exact equality, 
the system of $3 N_\nu$ + $3 N_{\bar \nu}$ 
differential equations in 
\eq\bref{1p7} 
reduce to six equations.  
One sums over $j$ and $k$ 
in \eq\bref{1p7}  
to obtain 
\ct{ks94b} 
\beq  
 {{d\vec r_{v}} \over {dt}} = 
    \vec r_{v} \times 
    \lrb{ {{\vec \De } \over {2 E_0}} - 
           {\vec V_{\nu \nu}} }
\ , \quad \quad 
 {{d\vec r_{w}} \over {dt}} = 
    \vec r_{w} \times 
    \lrb{ {{\vec \De } \over {2 \overline E_0}} + 
           {\vec V^{*}_{\nu \nu}}  } 
\quad ,  
\label{2p3} 
\eeq 
where 
\beq  
  \vec V_{\nu \nu} = \sqrt{2} G_F 
     \lrb{ n_{\nu} \vec r_v - n_{\bar \nu} \vec r_w^{\ *} } 
\quad ,     
\label{2p4} 
\eeq 
and 
$1/E_0$ and $ 1/ \overline E_0$ 
are  the average inverse neutrino and antineutrino energies  
given by 
\beq  
  \fr{1}{E_0} \equiv  
  \fr 1 {N_\nu}  \sum_j \fr{1}{E^j}
\ , \quad \quad 
  \fr{1}{\overline E_0} \equiv  
  \fr{1}{N_{\bar \nu}}  \sum_k \fr{1}{\overline E^k} 
\quad . 
\label{2p5} 
\eeq  

Initial flavor eigenstates correspond to  
\beq  
   \vec r_v (0) = \lrb{ 1, 0, 0 }  
\ , \quad 
   \vec r_w (0) = \lrb{ 1, 0, 0 }  
\quad . 
\label{2p6} 
\eeq  

\secttit{III.\ The Solution}  
\setcounter{section}{3}   
\setcounter{equation}{0}   

It is convenient to change variables.  
First, we use $\tau$ in lieu of $t$,  
where 
\beq  
 \tau \equiv  \taufac \ t 
\label{3p1} 
\eeq   
with  
\beq  
  \kappa_0 \equiv \kapo 
\quad . 
\label{3p2} 
\eeq   
Second, we use a mass eigenstate basis.  
This involves a rotation of vectors 
by an angle of $2 \theta$.  
We denote the new vectors 
for neutrinos by $\vec y$ 
and for antineutrinos by $\vec z$.  
They are given by 
$$ 
  y_1(\tau ) \equiv r_{v1}(t) \ctth - r_{v2}(t) \stth 
\ , \quad  
$$ 
$$ 
  y_2(\tau ) \equiv r_{v1}(t) \stth + r_{v2}(t) \ctth 
\ , \quad  
$$ 
\beq  
  y_3(\tau ) \equiv r_{v3}(t) 
\quad , 
\label{3p3} 
\eeq  
and 
$$ 
  z_1(\tau ) \equiv \nubarr \lrb{ r_{w1}(t) \ctth - r_{w2}(t) \stth }
\ , \quad  
$$ 
$$ 
  z_2(\tau ) \equiv \nubarr \lrb{ r_{w1}(t) \stth + r_{w2}(t) \ctth } 
\ , \quad 
$$ 
\beq  
  z_3(\tau ) \equiv  - \nubarr r_{w3}(t)  
\quad . 
\label{3p4} 
\eeq  
For antineutrinos, 
a factor of $\nubarrs$ is included for convenience,   
and the sign of the third component 
is flipped, which corresponds to 
using the ``$*$'' vector.  
Equation \bref{2p3}  
becomes 
\beq 
 \ddtau{\vec y} = 
   \vec y \times 
  \lrb{ \gam \hat e_1 - \fr{1}{\gam} \lrb{\vec y - \vec z} } 
\ , \quad \quad  
 \ddtau{\vec z} = 
   - \vec z \times 
  \lrb{  \gab \hat e_1 + \fr{1}{\gam} \lrb{\vec y - \vec z} } 
\quad , 
\label{3p5} 
\eeq  
where 
\beq 
  \gamma \equiv \sqrt{\kappa_0} 
\ , \quad 
  \bar \gamma \equiv \fr{E_0}{\overline E_0} \gamma
\quad . 
\label{3p6} 
\eeq 
Initial flavor eigenstates  
correspond to    
\beq  
   \vec y (0) = \lrb{ \ctth , \stth, 0 } 
\ , \quad 
   \vec z (0) = \nubarr \lrb{ \ctth , \stth, 0 } 
\quad . 
\label{3p7} 
\eeq

Equation \bref{3p5} 
is solved 
for arbitrary initial conditions 
in Appendix A.  
The solution for $y_1$ and $w_1$ 
is in terms of Jacobi elliptic functions.   
Thus the motion of the first components 
of $\vec y$ and $\vec w$ 
is period in $\tau$ 
with a period which we denote by ${\cal T}_f$.  
See \eqs\bref{ap18} and \bref{ap20}.  
The solutions for the second and third components are 
\beq  
 y_2 (\tau) =  \sqrt{1 - y_1^2 (\tau) } \ \cos \varphi (\tau)  
\ , \quad \quad 
 y_3 (\tau) =  \sqrt{1 - y_1^2 (\tau) } \ \sin \varphi (\tau)   
\quad ,
\label{3p8} 
\eeq  
$$   
 z_2 (\tau) =  
   \sqrt{\lrb{\nubarr}^2 - \lrb{\ka-y_1 (\tau) }^2 }  
      \ \cos \bar \varphi (\tau)   
\ , 
$$ 
\beq
 z_3 (\tau) =  
   \sqrt{\lrb{\nubarr}^2 - \lrb{\ka-y_1 (\tau) }^2 } 
      \ \sin \bar \varphi (\tau)
\quad , 
\label{3p9} 
\eeq  
where $k_1 = y_1 \lrb{0} - z_1 \lrb{0}$ and 
where $\varphi$ and $\bar \varphi$ 
are complicated functions 
of $y_1$.  
Because $y_1$ is periodic, 
$\cos \varphi (\tau)$ and $\sin \varphi (\tau) $ 
are periodic with a period ${\cal T}_s$.  
Likewise, 
$\cos \bar \varphi (\tau)$ and $\sin \bar \varphi (\tau) $ 
are periodic with a period $\bar {\cal T}_s$.  
See \eq\bref{ap25}.  
The periods in $t$ are related to those in $\tau$ 
by the factor 
$2{E_0}\sqrt{\kappa_0}/{\Delta}$ according to 
\eq\bref{3p1}.  

Hence, the motion of the second and third components 
is bimodal.  
They involve two periodic functions 
each with a different period.  
In terms of the original variables 
$\vec r_v$ and $\vec r_w$, 
all components are bimodal because 
$r_{v1}$ and $r_{w1}$ involve $y_2$ and $w_2$ 
when \eqs\bref{3p3} and \bref{3p4} 
are inverted.  
In general, 
the periods are not compatible 
and the curves for 
$r_{v1}$, $r_{v2}$, $r_{v3}$, 
$r_{w1}$, $r_{w2}$ and $r_{w3}$ 
appear irregular.  
When the gas has only neutrinos 
and no antineutrinos, 
the slow mode goes away 
and the behavior is purely periodic.  
This was the case observed in 
ref.\ \ct{samuel93a}.  
Likewise, 
in the CP-symmetric case, 
in which  
$N_\nu = N_{\bar \nu}$ and 
$E_0 = \bar E_0$, 
the periods ${\cal T}_s$ and $\bar {\cal T}_s$ 
go to infinity and the motion also becomes periodic 
with period ${\cal T}_f$.  
This is the case treated in
ref.\ \ct{ks94b}.  

\secttit{IV.\ Explicit Examples}  
\setcounter{section}{4}   
\setcounter{equation}{0}   

The analytic solution in Appendix A 
is quite complicated.  
To obtain insight, 
it is useful to consider 
specific cases.  
Below, the values for $\sttth$, $\Delta$, 
$n_{\nu}$, $n_{\bar \nu}$, etc., 
are not chosen to correspond to those 
in physical systems, 
such as in the early Universe,  
but to best illustrate different behaviors. 

It is convenient to use $\tau$ 
in \eq\bref{3p1} 
since then the system depends on only four parameters, 
namely,  
$\stth$, $\gamma$, $\bar \gamma$ and $\nubarrs$.  
Although solutions also depend on the initial values of 
$\vec y \lrb{0}$ and $\vec z \lrb{0}$, 
in this section, 
we avoid this dependence 
by using the initial flavor eigenstates 
in \eq\bref{3p7}.   
We select $\stth = 0.8$.  
The large value for the vacuum mixing angle  
amplifies effects.  

Figure 1 shows the three components of $\vec y \lrb{\tau}$ 
for the case 
$\nubarrs = 0.9$, $\gamma = 0.1$ and $\bar \gamma = 0.12$.  
Since the solution for $y_1 \lrb{\tau}$ is always 
in terms of a Jacobi elliptic function, 
the graph of $y_1 \lrb{\tau}$ is periodic. 
This is evident in Figure 1a. 
Since two periods are involved 
in $y_2 \lrb{\tau}$ and $y_3 \lrb{\tau}$, 
the corresponding curves are bimodal 
and irregular,  
as Figures 1b and 1c show.  
The plots for $z_1 \lrb{\tau}$, 
$z_2 \lrb{\tau}$ and $z_3 \lrb{\tau}$ 
are qualitatively similar to 
$y_1 \lrb{\tau}$, 
$y_2 \lrb{\tau}$ and $y_3 \lrb{\tau}$.  
In other words, 
$z_1 \lrb{\tau}$ is periodic,  
and $z_2 \lrb{\tau}$ and $z_3 \lrb{\tau}$ 
are bimodal and irregular.  
To save space, we do not display them.  
Additional insight is gained by examining the orbit 
for neutrinos, \ie, 
the trajectory 
$\lrb{y_1 \lrb{\tau},y_2 \lrb{\tau},y_3 \lrb{\tau}}$ 
in three-space.  
Figure 2a displays the neutrino orbit.  
The motion is as though one were moving back and forth 
over a cup-shaped surface 
in which the cup is also slowly rotating.  
These two motions crudely represent the two modes: 
The back-and-forth mode has a period of ${\cal T}_f$.   
The other mode corresponds to the rotation of the cup,  
is governed by the period ${\cal T}_s$, 
and   
is controlled by 
$\varphi \lrb{\tau}$.  
The antineutrino orbit,  
given by $\vec z \lrb{\tau}$, 
is similar to 
$\vec y \lrb{\tau}$
except that 
``little loops'' are made near the ``edge'' of the cup.  
See Figure 2b.  
In the flavor eigenstate basis, 
all components are bimodal and irregular.  
Figure 3 displays  
$
 r_{v1} \lrb{\tau} \equiv  
 \ctth \, y_1 \lrb{\tau} + \stth \, y_2 \lrb{\tau} 
$.  
Qualitatively $r_{w1} \lrb{\tau}$ 
is similar to $r_{v1} \lrb{\tau}$.    
The case in this paragraph 
is near the CP-symmetric limit for which 
$\nubarrs = 1.0$ and $\gamma = \bar \gamma $.  
In that limit, 
the motion is back-and-forth over the cup 
but the cup does not rotate 
\ct{ks94b} 
so that the motion is a single mode and periodic 
in all variables.  

When $\nubarrs$ is reduced, 
the amplitude for the back-and-forth motion 
becomes smaller and 
does not pass over the ``top'' of the cup.  
Figure 4a shows the orbit of $\vec y \lrb{\tau}$ 
for the case  
$\nubarrs = 0.75$, $\gamma = 0.1$ and $\bar \gamma = 0.12$.  
The antineutrino orbit is similar to the neutrino orbit 
and again has little loops. 
See Figure 4b.  
In terms of components, 
$y_1 \lrb{\tau}$ and $z_1 \lrb{\tau}$ have graphs 
similar to the one in Figure 1a but the amplitude is smaller.  
The graphs for $y_2 \lrb{\tau}$, $y_3 \lrb{\tau}$, 
$z_2 \lrb{\tau}$, $z_3 \lrb{\tau}$ 
look like cosine and sine functions with ``small bumps''.  
Figure 5, the plot of $y_2$ versus $\tau$,
exemplifies this.  

When $\nubarrs = 0$, 
one obtains the pure neutrino gas with no antineutrinos.  
The orbit for the neutrinos 
in this limiting case, 
which was analytically obtained 
in ref.\ \ct{ks94b}, 
is a circle.    
Figure 6a displays the neutrino orbit for the case 
$\nubarrs = 0.4$, $\gamma = 0.5$ and $ \gab = 0.6$.  
As expected, 
the orbit is approximately a circle.  
However, because $\nubarrs \ne 0$, 
there are ``wiggles''.  
The orbit looks somewhat like a ``lasso''.  
The antineutrino orbit is ``flowerlike''.  
See Figure 6b.  

It turns out that, in the above three examples,  
$\bar {\cal T}_s = {\cal T}_s$.  
For $\gam$ and $\gab$ not too large, 
this equality is general 
and holds for other values of the parameters 
$\stth$, $\nubarrs$, etc..  
Hence the angular motions,  
$\varphi \lrb{\tau}$ and $\bar \varphi \lrb{\tau}$,  
do not get out of phase. 
As a consequence, 
the initial CP-asymmetry 
is not amplified in time.  
For the discussion in the Section VI, 
we refer to this phenomenon as 
$\varphi$-{\it phase locking}.  

However, when $\gam$ and $\gab$ become sufficiently large,  
the equality $\bar {\cal T}_s = {\cal T}_s$ is suddenly violated.  
For initial flavor conditions, 
this happens at the point in parameter space when  
\beq 
 \nubarr \lrb{1-\ctth} = \xp 
\quad ,  
\label{4p1} 
\eeq 
where $\xp$ is given in 
\eq\bref{ap17}.  
This is the point at which the denominator factor 
$ 
 {{\lrb{ \nubarrs}}^2} - 
        {{\left(  { \ya} - { \ka} \right) }^2} 
$ 
in $\dphidsbar$ vanishes 
for the maximum value obtained by  $y_1$.  
See \eq\bref{ap23}. 
For $\gam$ or $\gab$ above the critical point,  
$\bar {\cal T}_s \ne {\cal T}_s$.  
In this case, 
even a small initial CP-asymmetry 
can eventually grow into a large CP-asymmetry 
at certain times.  

As an example of this phenomenon, consider the case with  
$\nubarrs = 0.9$, $\gamma = 0.5$ and $ \gab = 0.5$.  
One finds 
${\cal T}_f \approx 2.2832$, 
${\cal T}_s \approx 4.8383$, 
$\bar {\cal T}_s \approx 4.3235$.  
Figures 7a and 7b compare 
$y_2 \lrb{\tau}$ and $z_2 \lrb{\tau}$.  
By the time $\tau$ is $10$, 
$y_2 $ and $z_2 $ 
are about one-sixth of a period out of phase 
so that a sizeable CP-asymmetry is built up, 
even though the initial CP-asymmetry, 
as measured by the difference of $\nubarrs$ from one, 
is $10$ \%.  
The orbits for this case are fairly irregular.  
See Figures 8a and 8b.  

\secttit{V.\ Relation to Behaviors Observed in the Early Universe}  
\setcounter{section}{5}   
\setcounter{equation}{0}   

In a certain region of parameter space 
for $\Delta < 0$, 
ref.\ \ct{ks93a} 
observed irregular flavor behavior.  
It is interesting to ask whether this could be 
bimodal self-maintained coherence.  
The equations governing 
neutrino oscillations in the early Universe 
\ct{kps93a,ks94a} 
differ from those in 
\eq\bref{1p7}  
in two repects:   
First, there are CP-asymmetric effects. 
This means that the medium-induced linear term, 
the analog of $\vec \Delta$ 
in \eq\bref{1p8} 
is slightly different for neutrinos and antineutrinos. 
Second, 
$\vec B_v^j$ and $\vec B_w^k$ are time dependent 
due to the expansion of the Universe.  
However, 
in much of parameter space,  
interaction rates are faster than  
the expansion rate of the Universe, 
so that for short times 
$\vec B_v^j$ and $\vec B_w^k$ can be considered constant.  
Over long times, however, effects due to 
a non-constant Hubble constant can be important.  

The CP-asymmetric effects, 
which are due to the excess of  
electrons over positrons, 
are relatively small.  
However, they can eventually lead 
to a situation in which antineutrino vectors 
are not exactly aligned with neutrino vectors.  
Since the neutrino gas is extremely dense, 
small CP-asymmetries in the neutrino sector 
may lead to a sizeable interaction in 
$\vec V_{\nu \nu}$.  
Hence, the electron-positron asymmetry 
might be important indirectly 
via the nonlinear neutrino-neutrino term. 

Given this idea, 
it is reasonable to assume that 
the simplified model 
in Sect.\ II 
can represent the coherent behaviors 
for the early Universe,  
for the case where the energy distributions 
of neutrinos and antineutrinos are the same.  
The main effect of the positron-electron asymmetry 
is to misalign $\vec y$ and $\vec z$.  
Hence, it is of interest to examine solutions 
to \eq\bref{2p3}  
with $E_0 = \overline E_0$, $n_\nu = n_{\bar \nu}$ and 
$\vert \vec r_v \vert = \vert \vec r_w \vert$,  
but with $\vec r_v \lrb{0} \ne \vec r_w \lrb{0}$.  

We perform the same change of variables as 
in \eqs\bref{3p3} and \bref{3p4}.  
The CP-asymmetric effects are then represented 
as a change in the initial conditions for $\vec z$ via 
\beq 
  \vec z \lrb{0} = \nubarr 
    \lrb{ \cos \lrb{2\theta - 2\alpha},  
          \sin \lrb{2\theta - 2\alpha} \cos 2\beta, 
          \sin \lrb{2\theta - 2\alpha} \sin 2\beta} 
\quad , 
\label{5p1} 
\eeq 
while keeping $\vec y \lrb{0}$ as 
in \eq\bref{3p7},  
so that $\vec y \lrb{0} = \lrb{ \ctth , \stth, 0 } $. 
Here, $\alpha$ and $\beta$ are angles 
that control the CP-asymmetry 
in the neutrino sector 
in the $1$-$2$ plane and in the $3$ direction.  

Figure 9 displays the orbits 
for the neutrinos and antineutrinos 
for $\stth = 0.8$, $\sin 2\alpha = 0.01$, 
$\sin 2\beta = 0.0$,  
$\gam =\gab = 0.1$ 
and 
$\nubarrs = 1.0$.  
The motion is mostly back-and-forth 
as in the CP-symmetric case of 
reference \ct{ks94b} 
but the second mode causes the back-and-forth cycle 
to shift somewhat and not retrace over itself.  
When both $\alpha$ and $\beta$ are non-zero, 
fairly irregular behavior can arise.  
See Figure 10 
which displays the orbits 
for the neutrinos and antineutrinos 
for $\stth = 0.8$, $\sin 2\alpha = 0.1$, 
$\sin 2\beta = 0.1$, and 
$\gam =\gab = 0.1$. 
These examples show that 
bimodal self-maintained coherence 
and irregular flavor dependence 
arise even when the only CP-asymmetry 
is in the initial conditions 
for neutrinos and antineutrinos.  

We have examined some of the orbits 
in the simulations of 
refs.\ \ct{ks93a,ks95b} 
to see whether bimodal coherence occurs 
in the early Universe.  
Due to the variety of manifestations 
of bimodal solutions, 
as typified in Figures 1--10, 
it is not always so easy to be sure 
that the irregular behavior seen  
corresponds to bimodal coherence.  
However, 
many results from oscillations in the early Universe 
are similar 
to the examples considered here.  
One clear case 
from ref.\ \ct{ks95b} 
occurs for the simulations with 
$\Delta = - 1.0 \times 10^{-4}$ eV$^2$, 
$\sttth = 0.49$ at approximately $0.13$ seconds 
after the Big Bang.  
Figure 11 displays the orbit for the neutrinos 
in the flavor basis.  
The orbit for antineutrinos is virtually the same 
because of the CP-suppression mechanism 
discovered 
in refs.\ \ct{kps93a,ks94a}. 
Two modes are clearly seen.  
The orbit in Figure 11 is qualitatively similar 
to the orbit in Figure 2.   
Another example from 
ref.\ \ct{ks95b} 
occurs for $\Delta = - 1.0 \times 10^{-6}$ eV$^2$, 
$\sttth = 1.0 \times 10^{-8}$ 
at approximately $0.94$ seconds. 
Although we do not display the orbit, 
it is similar to Figure 9a.  

Bimodal self-maintained coherence 
might be the general behavior 
of many dense neutrino gas systems.  
All three behaviors observed in the simulations 
of refs.\ \ct{ks94a,ks93a,ks95b} 
are qualitatively obtained as limits. 
Irregular behavior corresponds to bimodal coherence 
when both modes have sizeable amplitudes.  
Residual self-maintained coherence occurs 
when only one mode has a sizeable amplitude.  
Finally, smooth behavior arises when vectors point 
in the directions of the 
the linear terms in the effective magnetic fields,   
so that both modes have small amplitudes.  
Indeed, 
for a dense neutrino gas in the early Universe,  
tiny rapid fluctuations were observed 
in the simulations of 
refs.\ \ct{kps93a,ks94a,ks93a,ks95a,ks95b} 
in the smooth region of the phase diagrams.  
One possible explanation for them is that 
they correspond to small-amplitude bimodal coherence. 
Of course, when the amplitudes for oscillations are tiny,  
other effects may create them.  
Even if all the behaviors in parameter space 
for $\Delta < 0$ correspond to bimodal coherence, 
it is still useful to characterize them by irregular, 
residual self-maintained coherence, and smooth 
as in Figure 1 
of ref.\ \ct{ks93a}.

\secttit{VI.\ Neutrino Behavior in the Early Universe for $\mu_\nu \ne 0$}  
\setcounter{section}{6}   
\setcounter{equation}{0}   

The studies of neutrino oscillations in the early Universe 
in refs.\ \ct{kps93a,ks94a,ks93a,ks95a,ks95b} 
were performed under the assumption that $n_\nu = n_{\bar \nu}$ 
and that the energy distributions for neutrinos and for antineutrinos 
are the same.   
Given that there is a CP-asymmetry for charged leptons, 
namely the excess of electrons over positrons, 
it is possible that 
the numbers of neutrinos and antineutrinos 
are not equal,  
in which case a chemical potential potential $\mu_\nu$  
for neutrinos needs to be introduced.  

One purpose of the current work is to gain 
insight into the asymmetric neutrino-antineutrino gas.  
With this insight, 
it is possible to speculate on the phase diagram 
and on the flavor behavior.     
To rigorously determine these, 
simulations must be performed, 
but such computer studies are numerically 
intensive because one must 
explore a phase space of three parameters: 
$\theta$, $\Delta$ and $\mu_\nu$. 
A thorough study cannot be undertaken 
with the power of present computers.  
However, an educated guess as to the results can be made.  

Due to the explicit asymmetry 
between neutrinos and antineutrinos 
when $\mu_\nu \ne 0$,  
residual self-maintained coherence, 
which corresponds to a situation 
in which only a single mode is excited, 
is unlikely to appear, 
except for perhaps brief periods  
for exceptional values of parameters.  
Instead, irregular flavor, as governed by bimodal coherence,  
should replace it.  
Thus the diagram for $\Delta < 0$  
probably has only two phases: smooth and irregular.  
When $\mu_\nu \ne 0$, 
irregular behavior should arise 
in the regions of the phase diagrams 
of refs.\ \ct{ks93a,ks95b}
where irregular or residual self-maintained coherence 
occurred.  

For $\Delta > 0$, 
it is likely that both neutrino and antineutrino vectors 
follow their respectively effective magnetic fields, 
thereby maintaining themselves in 
approximate nonlinear mass eigenstates.  
Hence, smooth behavior is generally  expected 
for $\Delta > 0$.  
However, if $\overline E_0$ is significantly 
different from $E_0$ then 
transitory weak irregular behavior should take place 
in the ``cross-over region''.  
The cross-over region is the period during which   
neutrinos evolve  
from approximate flavor eigenstates 
to approximate vacuum mass eigenstates 
\ct{ks94a}. 
Small amplitude irregular behavior should 
appear there because neutrinos and antineutrinos 
make the transition to mass eigenstates 
at different times. 
This cause neutrino and antineutrinos vectors 
to point in different directions. 
The effect should be most pronounced for small $\Delta$, 
particularly 
when $\Delta < 10^{-9}$ eV$^2$. 
After the transition region, 
the behavior should become smooth again. 
A similar transitory effect 
might also occur in the smooth phase 
for $\Delta < 0$.     

One motivation for considering 
a non-zero neutrino chemical potential  
is that 
it introduces additional CP-asymmetry 
into the system.  
Reference \ct{lpst87a} 
showed that a large neutrino-antineutrino asymmetry 
can have an effect on nucleosynthesis.  
However, 
for the values of the parameters of the early Universe, 
$\varphi$-phase locking is expected.  
This implied that the CP-asymmetry 
due to  $\mu_\nu \ne 0$ 
should not be amplified in time.  
Hence, 
neutrino oscillations for $\mu_\nu \ne 0$ 
are likely to affect nucleosynthesis 
at the same level as 
in the case of no neutrino masses and mixing 
\ct{dt92a,df92a}.  
Constraints on the chemical potential of neutrinos 
from Big Bang nucleosynthesis have been obtained 
in ref.\ \ct{smf91a}.  

\secttit{VII.\ Summary}  
\setcounter{section}{7}   
\setcounter{equation}{0}   

In this work, 
we analyzed neutrino oscillations 
in a dense gas 
in which there is an explicit neutrino-antineutrino asymmetry.  
The energy distribution for antineutrinos 
was not necessarily the same 
as the energy distribution for neutrinos, 
nor did the numbers of antineutrinos and neutrinos 
have to be the same.  

In the dense gas limit, 
alignment is expected to hold.  
Using this property, 
the non-linear multi-equation system 
was simplified to six equations 
in Section II.  
Remarkably, 
the six nonlinear equations were solvable.  
The solution was still quite complicated.   
After performing some transformations 
in Section III,  
most solution details were relegated  
to Appendix A.  
The generic behavior was bimodal self-maintained coherence.  
This is a collective soliton-like solution  
in which neutrino behavior is cooperative and 
in which two modes of different frequencies are present.  

Bimodal self-maintained coherence 
exhibits a wide range of behaviors 
depending on parameters and initial conditions.  
Particular illustrative cases were presented 
in Section IV.  

When the amplitude for one of the modes vanishes, 
ordinary self-maintained coherence is obtained.  
Such behavior was first observed in test simulations 
in ref.\ \ct{samuel93a}.  
It was subsequently observed in numerical simulations of
the early Universe  
in ref.\ \ct{ks93a} 
and understood analytically 
in ref.\ \ct{ks94b}.  
Another neutrino flavor behavior,  
which has been seen in the early Universe, 
is smooth 
\ct{kps93a,ks94a}.  
It corresponds to a bimodal solution in which the amplitudes 
of both modes are small.  
Finally, 
irregular behavior, 
which can occur for an inverted mass hierarchy 
\ct{ks93a},  
is mimicked 
by bimodal coherence 
when both amplitudes are present.  
These results suggest that bimodal coherence 
may be the general flavor behavior 
for many dense neutrino gases.  
As discussed in Section V, 
our results provide new insight 
into the simulations of neutrino oscillations 
in the early Universe 
performed 
in refs.\ \ct{kps93a,ks94a,ks93a,ks95a,ks95b}.  

With this additional understanding, 
it was possible to surmise 
the neutrino flavor dependence 
in the early Universe 
when neutrinos have a chemical potential, 
so that there is 
an explicit neutrino-antineutrino asymmetry.  
The expected behavior for $\Delta >0$ is smooth  
with a possible weak bimodal coherent phase 
for a short time period.  
The expected behaviors for the inverted mass hierarchy case, 
$\Delta <0$, 
are smooth and bimodal coherence.  
Bimodal coherence probably arises 
where irregular or 
residual self-maintained coherence 
arose for the $\mu_\nu =0$ case 
of refs.\ \ct{ks93a,ks95b}.  
Because of the phenomenon of $\varphi$-phase locking, 
effects on nucleosynthesis are not expected 
to be amplified by neutrino oscillations, 
as one might have a priori thought.  
The effects are probably limited to 
those directly due to a non-zero neutrino chemical potential 
\ct{smf91a}.  

In short, 
our work has shown that dense neutrino gases 
can exhibit interesting and non-intuitive phenomenon.  
The gases often have collective, 
cooperative and coherent flavor behavior.

\noindent 
\secttit{Acknowledgements} 

I would like to thank Alan Kosteleck\'y 
for discussions.  
I also thank Julius Wess for hospitality.  
This work is supported in part  
by the United States Department of Energy 
(grant number DE-FG02-92ER40698), 
by the Alexander von Humboldt Foundation, 
and by the PSC Board of Higher Education at CUNY. 

\secttit{Appendix A} 
\label{s:a} 
\setcounter{section}{8}   
\def\theequation{A.\arabic{equation}}
\setcounter{equation}{0}   

It turns out that \eq\bref{3p5} 
can be solved for arbitrary initial conditions.  
Four conserved quantities play a role.  
The first two 
\beq  
 { \ya}^2 + { \yb}^2 + { \yc}^2 = 1 
\label{ap1} 
\eeq  
and 
\beq  
 { \za}^2 + { \zb}^2 + { \zc}^2 = \lrb{\nubarr}^2 
\label{ap2} 
\eeq  
correspond to conservation of neutrinos 
and of antineutrinos.  
Equation \bref{3p5}  
also implies that 
\beq  
  \ya - \za = \ka 
\quad ,  
\label{ap3} 
\eeq  
and that 
\beq  
  y_1 - \fr{1}{2\gam\lrb{\gab+\gam}} 
   \lrb{ \lrb{y_2 - z_2}^2 + \lrb{y_3 - z_3}^2} = 
     \kab 
\quad , 
\label{ap4} 
\eeq  
where $\ka$ and $\kab$ are time-independent constants.  
They are determined by initial conditions, \ie, 
$$ 
  \ka = y_1 (0) - z_1 (0) 
\quad , 
$$ 
\beq   
  \kab = y_1 (0) - \fr{1}{2\gam\lrb{\gab+\gam}} 
   \lrb{ \lrb{y_2 (0) - z_2 (0) }^2 + 
         \lrb{y_3 (0) - z_3 (0) }^2} 
\quad . 
\label{ap5} 
\eeq  
For initial flavor eigenstates, 
$$ 
 {{{ \ka}}\rightarrow {{ \cs} \left( 1 - { \nubarr} \right) }} 
$$ 
\beq   
  {{{ \kab}}\rightarrow 
    {{ \cs} - {{\sttth}\over 
        {2 { \gam} \left( { \gab} + { \gam} \right) }}}} 
  {\left( 1 - { \nubarr} \right) }^2 
\quad . 
\label{ap6} 
\eeq

It turns out to be useful 
to solve for the quantity 
$\yb \zb + \yc \zc$ 
using 
\eqs\bref{ap1}--\bref{ap4}.  
One finds 
\beq   
  \yb \zb + \yc \zc = 
\factora  + 
  { \gam} \left( { \gab} + { \gam} \right)  
   \left( { \kab} - { \ya} \right)  + { \ka} { \ya} - 
  { \ya}^2  
\ . 
\label{ap7} 
\eeq

Thus, 
\eqs \bref{ap1}--\bref{ap4} 
allow one to eliminate 
four combinations of variables 
in favor of $y_1$: 
$$ 
 {\yb}^2 + { \yc}^2 \rightarrow 1 - { \ya}^2 
\quad , 
$$ 
$$ 
  \za \rightarrow   \ya - \ka  
\quad , 
$$ 
$$ 
{{{{{ \zb}}^2} + {{{ \zc}}^2}} \rightarrow 
   {{{\lrb{ \nubarr}}^2} - {{\left({ \ya} - { \ka} \right) }^2}}} 
\quad , 
$$ 
\beq  
{{{ \yb} { \zb} + { \yc} { \zc}}\rightarrow 
   {\factora + 
     { \gam} \left( { \gab} + { \gam} \right)  
      \left( { \kab} - { \ya} \right)  + { \ka} { \ya} - 
     {{{ \ya}}^2}}} 
\ . 
\label{ap8} 
\eeq  

Take the first order differential equation 
for $y_1$ in 
\eq\bref{3p5}, 
differentiate it with respect to $\tau$ 
and use the equations 
for the second and third components 
in \eq\bref{3p5}. 
It is a miracle that precisely 
the four quantities on the left-hand side of 
\eq\bref{ap8}  
enter.  
Straightforward but lengthy algebra gives 
$$ 
\dya2ds2 = 
\left( 1 + {{{ \gab}}\over {{ \gam}}} \right)  
   \left( { \yb} { \zb} + { \yc} { \zc} \right)  + 
$$ 
$$ 
  {{
   \left( { \ya} + { \za} \right)  
       \left( { \yb} { \zb} + { \yc} { \zc} \right)  - 
      { \ya} \left( {{{ \zb}}^2} + {{{ \zc}}^2} \right) 
  -\left( {{{ \yb}}^2} + {{{ \yc}}^2}  \right)  { \za}  }\over 
    {{{{ \gam}}^2}}}
$$ 
\beq  
 = -3 \Acoeffh y_1^2 - 2 \Bcoeffh y_1 - \Ccoeffh
\quad ,  
\label{ap9} 
\eeq  
where, in the last equality,  
the substitutions in \eq\bref{ap8}  
are used.  
The constants 
$\Acoeffh$, $\Bcoeffh$ and $\Ccoeffh$ 
are given by  
$$ 
 \Acoeffh = 
 {{{ \gab} + { \gam}}\over {{ \gam}}} 
\quad , 
$$ 
$$ 
 \Bcoeffh = 
{{{{\left( { \gab} + { \gam} \right) }^2}}\over 2} + 
  {{{{{ \ka}}^2}}\over {2 {{{ \gam}}^2}}} - 
  {{\left( { \gab} + { \gam} \right)  
      \left( { \ka} + { \kab} \right) }\over {{ \gam}}} 
\quad , 
$$ 
$$ 
 \Ccoeffh = 
   - {{\left( {\gab} + {\gam} \right) }^2} {\kab} - \factorb
$$ 
\beq  
   + {{\left( { \gab} + { \gam} \right)  
       }\over 
    {{ \gam}}} 
  \left({ \ka} { \kab} -   \factora  \right) 
\quad . 
\label{ap10} 
\eeq  
Thus the motion of $y_1$ 
corresponds to a classical particle moving 
in a one-dimensional cubic potential.  
This system can be solved by quadratures.  
Multiply \eq\bref{ap9} by $\ddtau{y_1}$ and integrate 
to obtain 
\beq  
  \half {\lrb{ \ddtau{y_1}}}^2 = - V \lrb{y_1} 
\quad ,  
\label{ap11} 
\eeq   
where 
\beq  
 V \lrb{\var} = 
 {- \half {{{\lrb{ \dyads (0) }}^2}}} + 
  { \Ccoeffh} \left( { \var} - { \ya (0) } \right)  + 
  { \Bcoeffh} \left( {{{ \var}}^2} - {{\lrb{ \ya (0) }}^2} \right)  + 
  { \Acoeffh} \left( {{{ \var}}^3} - {{\lrb{ \ya (0) }}^3} \right) 
\ . 
\label{ap12} 
\eeq   
The potential $V \lrb{\var}$ has three zeroes 
and is negative between the two larger zeroes.  
The motion of $y_1$ is between these two larger zeroes 
where $V \lrb{\var} \le 0 $.  
Integration of 
\eq\bref{ap11} leads to 
\beq  
 \tau = \int_{ y_1 (0)}^{y_1 (\tau)} \fr{dy}{\sqrt{-2 V(y)}}
\quad . 
\label{ap13} 
\eeq 
This equation gives $\tau$ in terms of 
$y_1 (\tau)$ 
where the latter appears as the upper limit of integration.  
Equation \bref{ap13} is inverted to obtain 
$y_1$ as a function of $\tau$.  

Since the potential is cubic,  
the solution is always related 
to Jacobi elliptic functions.  
In general, a cubic equation must be solved, 
so that the explicit solution is rather complicated. 
However, 
when 
\beq  
  \ddtau{y_1} (0) = 0 
\quad , 
\label{ap14} 
\eeq 
one can avoid this difficulty.  
The initial condition in \eq\bref{ap14} 
is an important case because it 
includes the initial flavor eigenstates 
in \eq\bref{3p7}.  
When 
\eq\bref{ap14} 
holds,  
the potential becomes%
\footnote{ 
For the case of general initial conditions 
one must shift $y_1$ by a constant 
which renders the middle zero at $0$.  
Then one proceeds as in the special case 
explicitly carried out here.} 
\beq  
  V \lrb{\var + {y_1} (0) } = 
    \var \lrb{\Acoeff \var^2 + \Bcoeff \var + \Ccoeff} = 
  \fr{\gab+\gam}{\gam} \var \lrb{\var - \xp} \lrb{\var + \xm}
\quad . 
\label{ap15} 
\eeq 
where 
$$ 
\Acoeff =\Acoeffh 
\quad , 
$$ 
$$ 
 \Bcoeff = \Bcoeffh + 3 \Acoeffh \ya (0) 
\quad , 
$$ 
\beq  
   \Ccoeff = 
   \Ccoeffh + 2 { \Bcoeffh} { \ya (0) } + 
              3 { \Acoeffh} {\lrb{{ \ya (0) }}^2} 
\quad . 
\label{ap16} 
\eeq  

The zeroes of the potential are 
${y_1} (0) - \xm$, ${y_1} (0)$ and ${y_1} (0) + \xp$ 
where 
\beq  
 \xp = {{- \Bcoeff + \sqr}\over {2 \Acoeff}} 
\ , \quad \quad 
 \xm = {{  \Bcoeff + \sqr}\over {2 \Acoeff}} 
\quad . 
\label{ap17} 
\eeq  
Thus $y_1$ oscillates between 
$y_1 (0)$ and ${y_1} (0) + \xp$.  

The integral in \eq\bref{ap13} is 
\beq  
\yasol (\s ) = 
{ \ya (0) } + { {{ \xm} { \xp}} \over {{ \xo}} } 
  {{ 
      {{{ \JacobiSN}^2(\s \sqrt{\fr{\xo \lrb{\gab+\gam}}{2 \gam}} ,
          {{\xp}\over {\xo}}) }}}\over 
    { { 
      {{{ \JacobiDN}^2(\s \sqrt{\fr{\xo \lrb{\gab+\gam}}{2 \gam}} ,
          {{{ \xp}}\over {{ \xo}}})}}} }} 
\quad ,  
\label{ap18} 
\eeq  
where 
\beq  
 \xo \equiv {{ \sqr} \over { \Acoeff}} = \xp + \xm  
\quad ,  
\label{ap19} 
\eeq  
and where $\JacobiSN$ and $\JacobiDN$ are 
the sine-amplitude and delta-amplitude 
Jacobi elliptic functions.  

Hence $y_1$ undergoes periodic motion with a period 
${\cal T}_f$ 
in $\tau$ given by 
\beq  
 {\cal T}_f = 
  2 \sqrt{ \fr{2 \gam}{\xm \lrb{\gab + \gam}}} 
  \ \EllipticK \lrb{- \qs} 
\quad ,  
\label{ap20} 
\eeq  
where 
\beq 
\EllipticK \lrb{- \qs} = 
  \int_0^1 { {dw} 
  \over 
    {\sqrt{\left( {1 -     w^2} \right) 
           \left( {1 + q^2 w^2} \right)}  }  } 
\quad ,  
\label{ap21} 
\eeq 
with $\qs \equiv {{ \xp} / { \xm}}$.  
The period $T_f$ in $t$ is 
\beq  
  T_f = \invtaufac \ {\cal T}_f 
\quad . 
\label{ap22} 
\eeq    

The solution for $z_1 (\tau)$ 
is given by using 
\eqs\bref{ap3} and \bref{ap18}.  

Finding  
the solution of 
the second and third components 
of $\vec y$ and $\vec z$ 
is not so straightforward.  
One proceeds by using the polar coordinate 
angles in the two--three plane, \ie, 
let $ \varphi = \tan^{-1} \lrb{ {y_3}/{y_2} } $ 
and  
$\bar \varphi = \tan^{-1} \lrb{ {z_3}/{z_2} } $. 
Differentiate $\varphi$ with respect to $\tau$ 
and 
use \eq\bref{3p5}.  
Do the same for $\bar \varphi$.  
Miraculously, 
only the four quantities 
in \eq\bref{ap8} enter.  
Hence, 
the substitutions in \eq\bref{ap8} 
can be used to express 
$\dphids$ and $\dphidsbar$ 
in terms of $y_1 (\tau)$.  
After straightforward but lengthy algebra,  
one finds  
$$ 
 \dphids = 
-{ \gam} + {{{ \ka} + \left(  
         { \gam} \left( { \gab} + { \gam} \right)  
          \left( { \kab} - { \ya} \right) \factorc \right)  { \ya}}\over 
    {{ \gam} \left( 1 - {{{ \ya}}^2} \right) }} 
\quad , 
$$ 
\beq 
 \dphidsbar = 
{ \gab} + {{\left( \factora + 
 { \gam} \left( { \gab} + { \gam} \right)  
          \left( { \kab} - { \ya} \right)  \right)  
       \left( { \ka} - { \ya} \right)  + 
      {{\lrb{ \nubarr}}^2} { \ya}}\over 
    {{ \gam} \left( {{\lrb{ \nubarr}}^2} - 
        {{\left(  { \ya} - { \ka} \right) }^2} \right) }} 
\quad .
\label{ap23} 
\eeq  
Thus, 
\beq  
     \varphi (\tau ' ) = 
      \varphi ( 0 ) + \int_0^{\tau'} d\tau \dphids  
\ , \quad \quad  
 \bar \varphi (\tau ' ) = 
 \bar \varphi ( 0 ) + \int_0^{\tau'} d\tau \dphidsbar  
\quad ,  
\label{ap24} 
\eeq 
where $\dphids$ and $\dphidsbar$ 
are given 
in \eq\bref{ap23} 
and where 
$$ 
 \varphi ( 0 ) = \tan^{-1} \lrb{\fr{y_3 (0)}{y_2 (0)}}
\ , \quad \quad 
 \bar \varphi ( 0 ) = \tan^{-1} \lrb{\fr{z_3 (0)}{z_2 (0)}} 
\quad . 
$$ 
The solutions for the second and third components 
are then given via 
\eqs\bref{3p8} and \bref{3p9}.  

Since $y_1$ is periodic, 
the angular motion in the two--three plane 
is also periodic but with a period 
different from ${\cal T}_f$.  
For neutrinos, 
let ${\cal T}_s$ denote the period in $\tau$.  
For antineutrinos, 
let $\bar {\cal T}_s$ be the period.  
Then, 
\beq 
            {\cal T}_s = 
  \fr{2 \pi {\cal T}_f }{\vert \int_0^{{\cal T}_f} 
         d\tau \dphids \vert} 
\ , \quad \quad  
  \overline {\cal T}_s = 
  \fr{2 \pi {\cal T}_f }{\vert \int_0^{{\cal T}_f} 
         d\tau \dphidsbar \vert} 
\quad ,  
\label{ap25} 
\eeq  
where $\vert \ \vert$ indicates absolute value.  
The subscripts $f$ and $s$ 
on ${\cal T}_f$ and ${\cal T}_s$
stands for ``fast'' and ``slow''.  
In the extreme dense gas limit ${\cal T}_f$ 
is much smaller than ${\cal T}_s$, \ie, 
${\cal T}_f$ is the period for the faster motion.  
The periods $T_s$ and $\overline T_s$ in terms of $t$ are 
\beq  
           T_s = \invtaufac \ {\cal T}_s
\ , \quad \quad 
 \overline T_s = \invtaufac \ \overline {\cal T}_s
\quad . 
\label{ap26} 
\eeq  

\vglue 0.6cm  
{\bf\large\noindent References}  
\vglue 0.4cm


\vglue 0.6cm  
{\bf\large\noindent Figure Captions}  
\vglue 0.4cm  

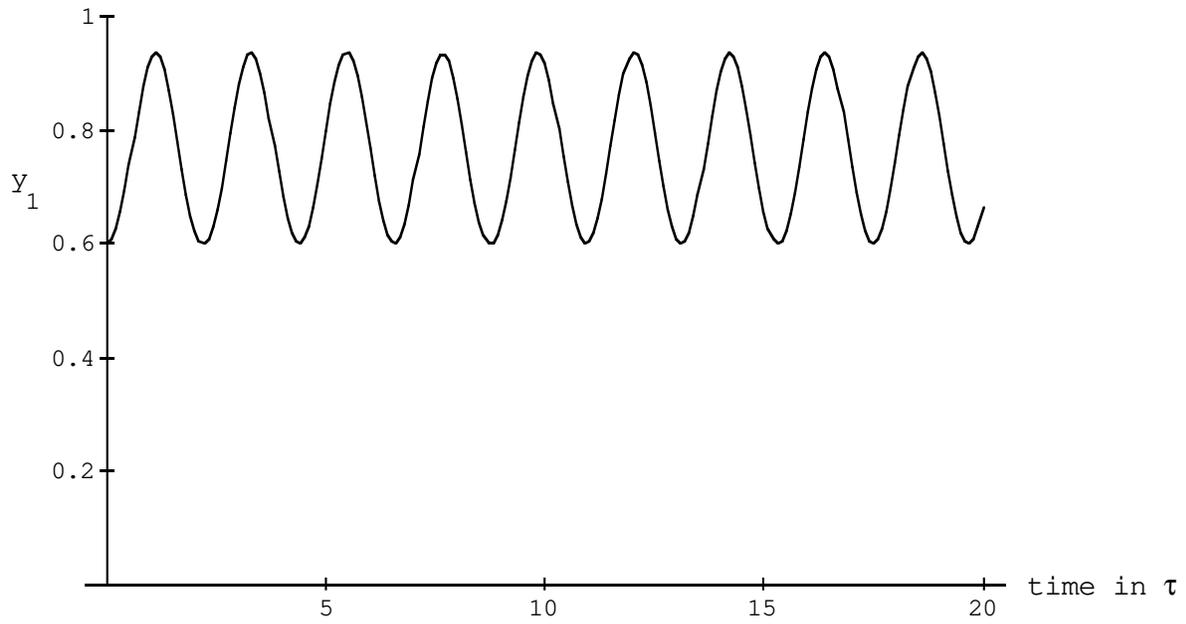
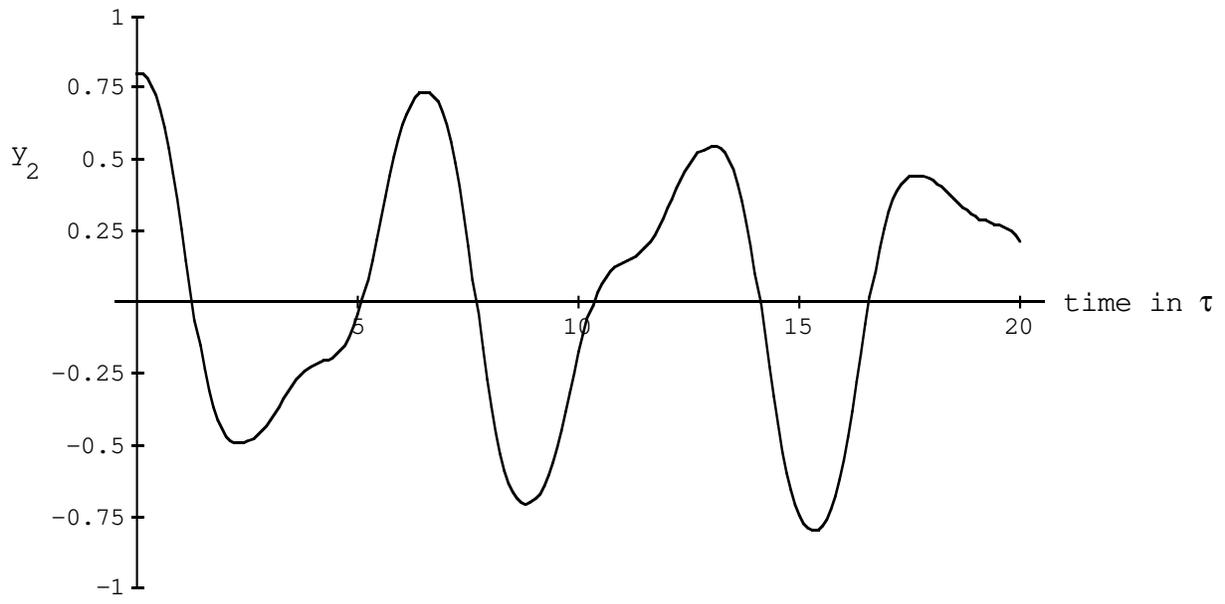
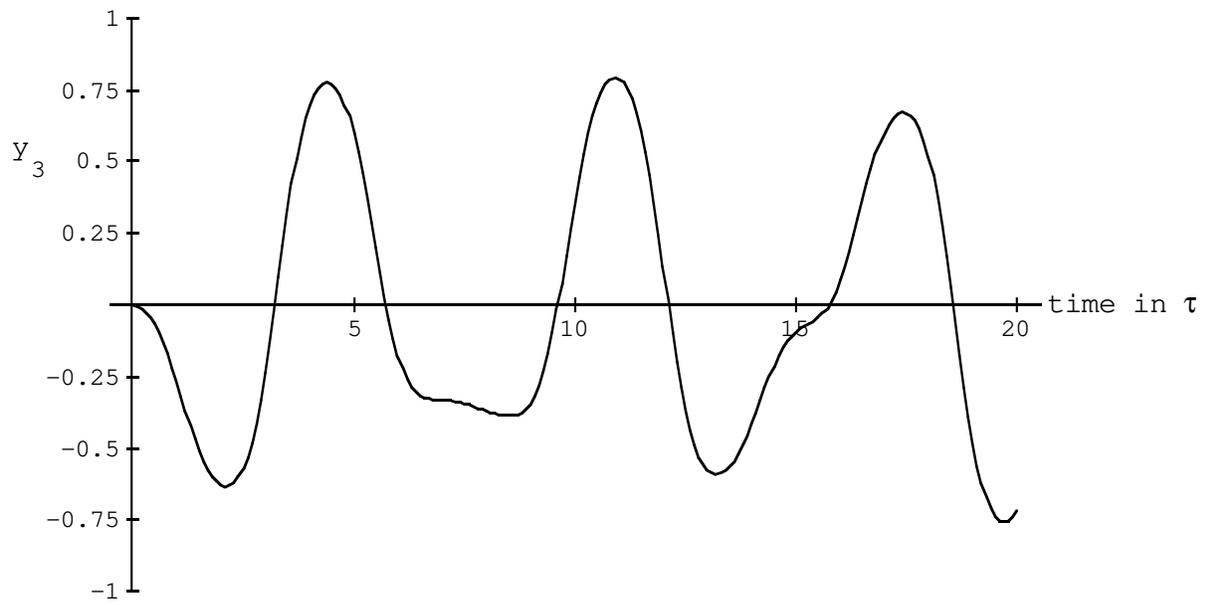
Figure 1. Components of $\vec y$
as a function of scaled time $\tau$ 
for $0 < \tau < 20$   
for the case 
$\stth = 0.8$, $\nubarrs = 0.9$, 
$\gamma = 0.1$ and $\bar \gamma = 0.12$ 
and with initial conditions as in \eq\bref{3p7}.
(a) The component $y_1$.
(b) The component $y_2$.
(c) The component $y_3$.

\medskip

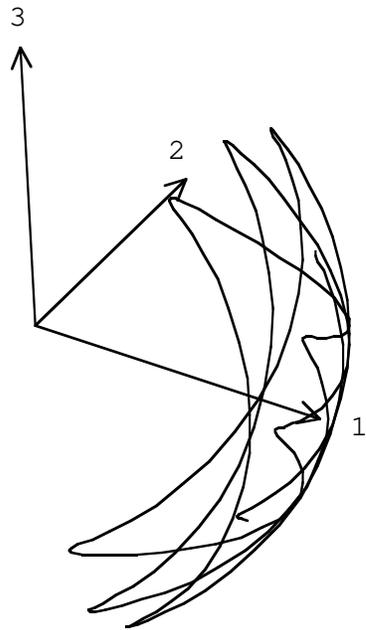
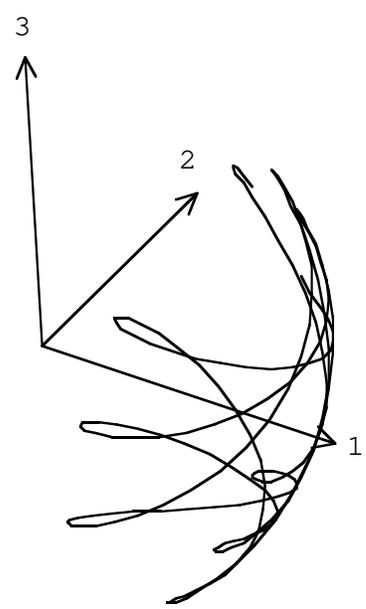
Figure 2. The three-dimensional orbits 
for the example in Figure 1. 
(a) The neutrino orbit. 
(b) The antineutrino orbit.  

\medskip

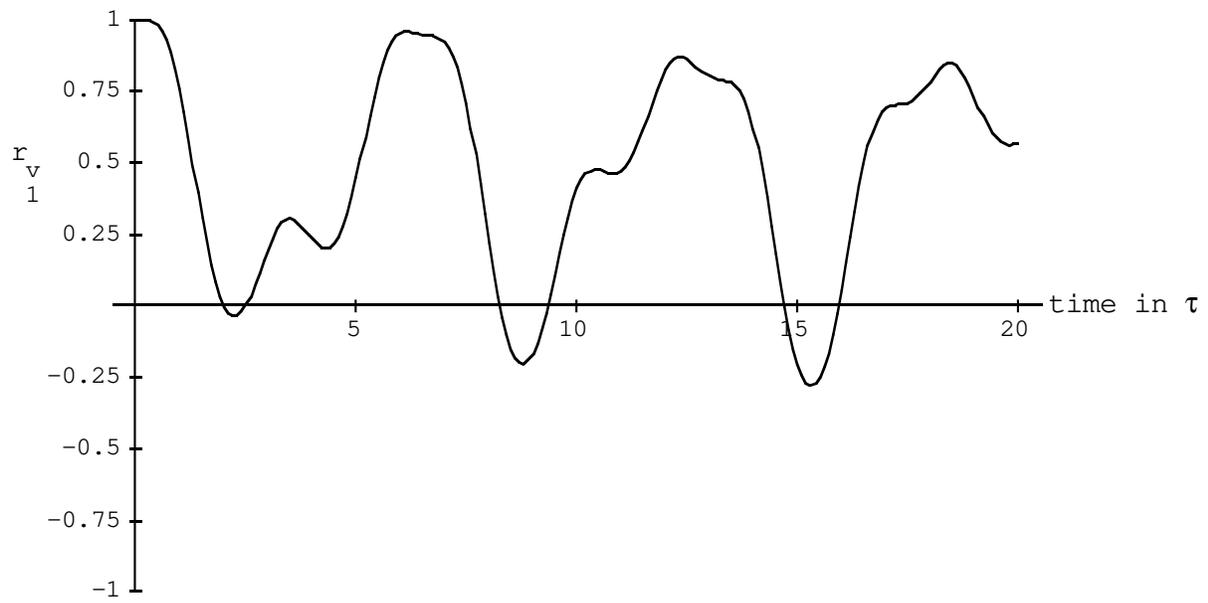
Figure 3. The flavor component $r_{v1}$ 
as a function of $\tau$.  
Parameters are the same as in Figure 1. 

\medskip

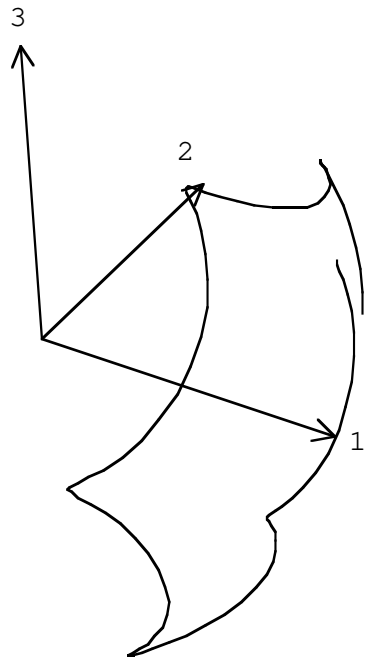
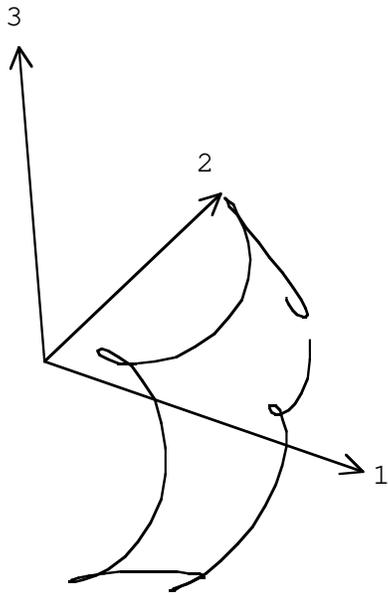
Figure 4. The orbits during 
$0 < \tau < 10$ 
for the case     
$\stth = 0.8$, $\nubarrs = 0.75$, 
$\gamma = 0.1$ and $\bar \gamma = 0.12$ 
and with initial conditions as in \eq\bref{3p7}.
(a) The neutrino orbit. 
(b) The antineutrino orbit.  

\medskip

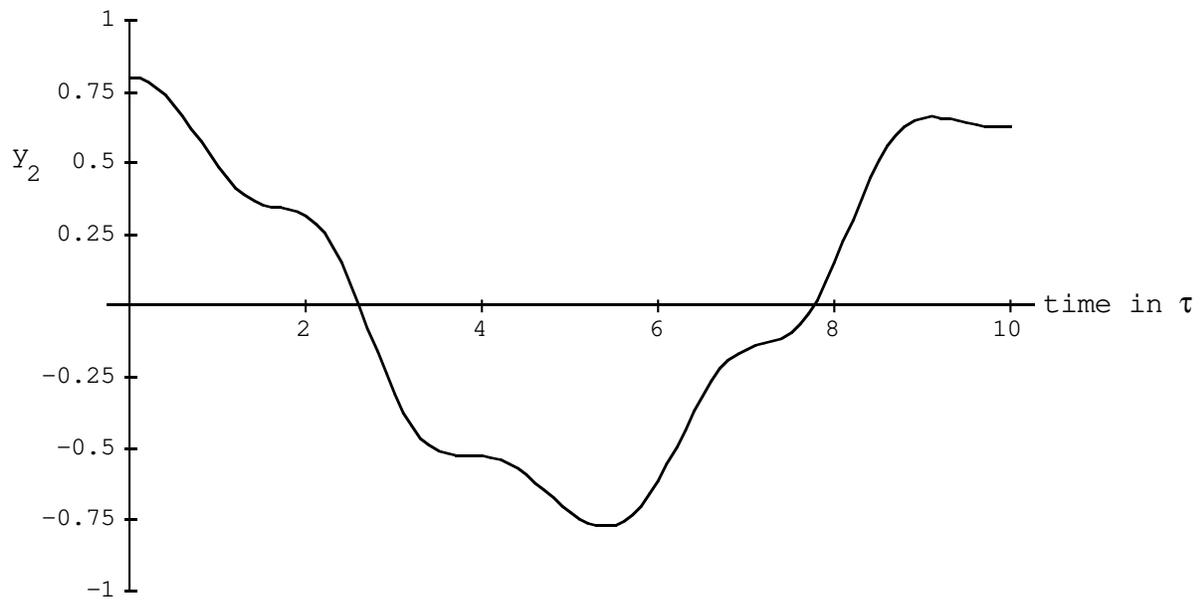
Figure 5. A plot of $y_2$ versus $\tau$ 
for the example in Figure 4.  

\medskip

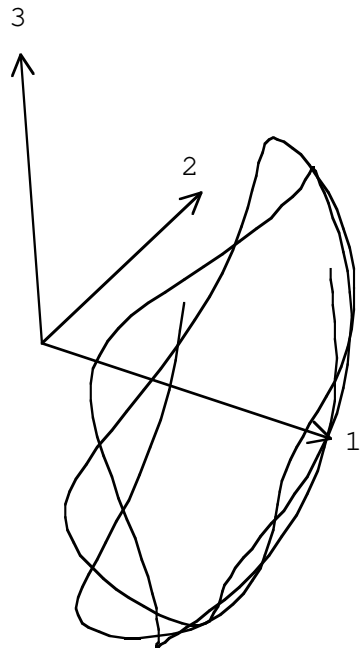
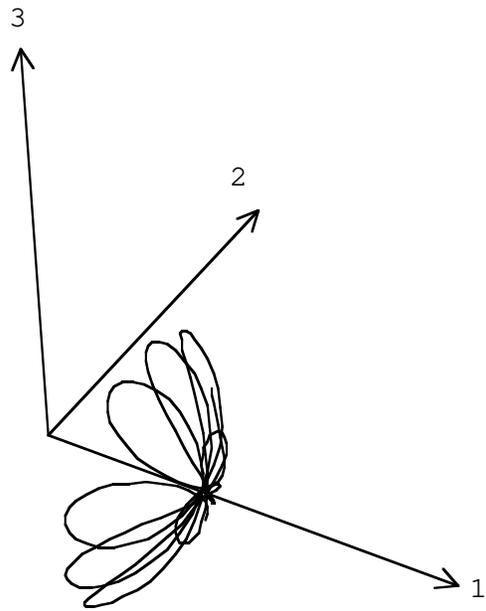
Figure 6. The orbits during 
for $0 < \tau < 20$ 
for the case     
$\stth = 0.8$, $\nubarrs = 0.4$, 
$\gamma = 0.5$ and $\bar \gamma = 0.6$ 
and with initial conditions as in \eq\bref{3p7}.
(a) The neutrino orbit. 
(b) The antineutrino orbit. 

\medskip

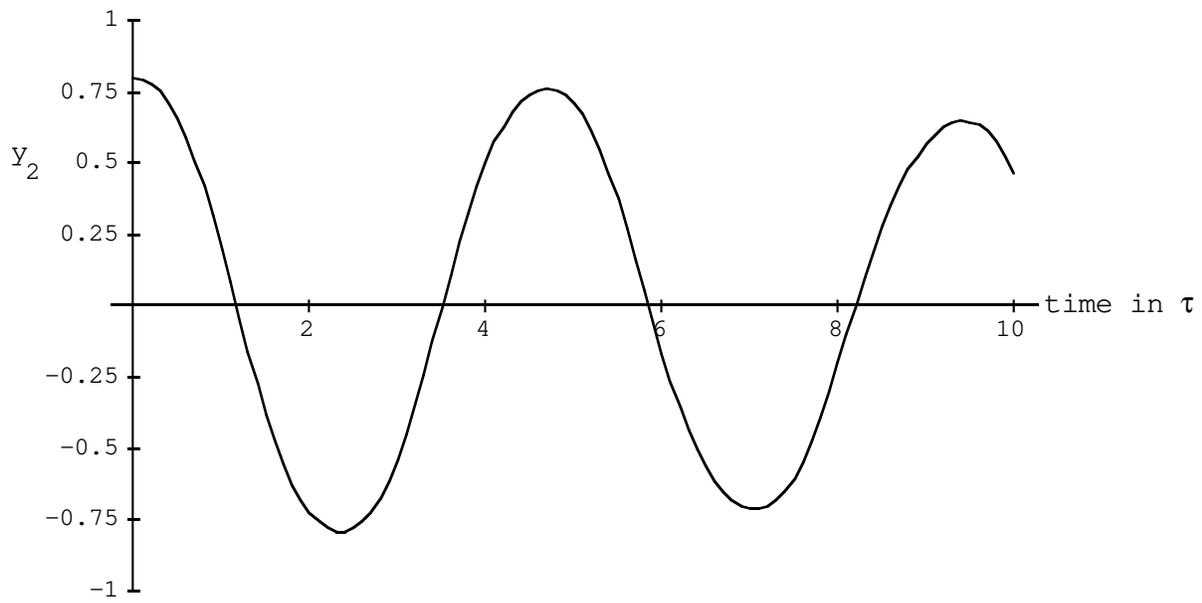
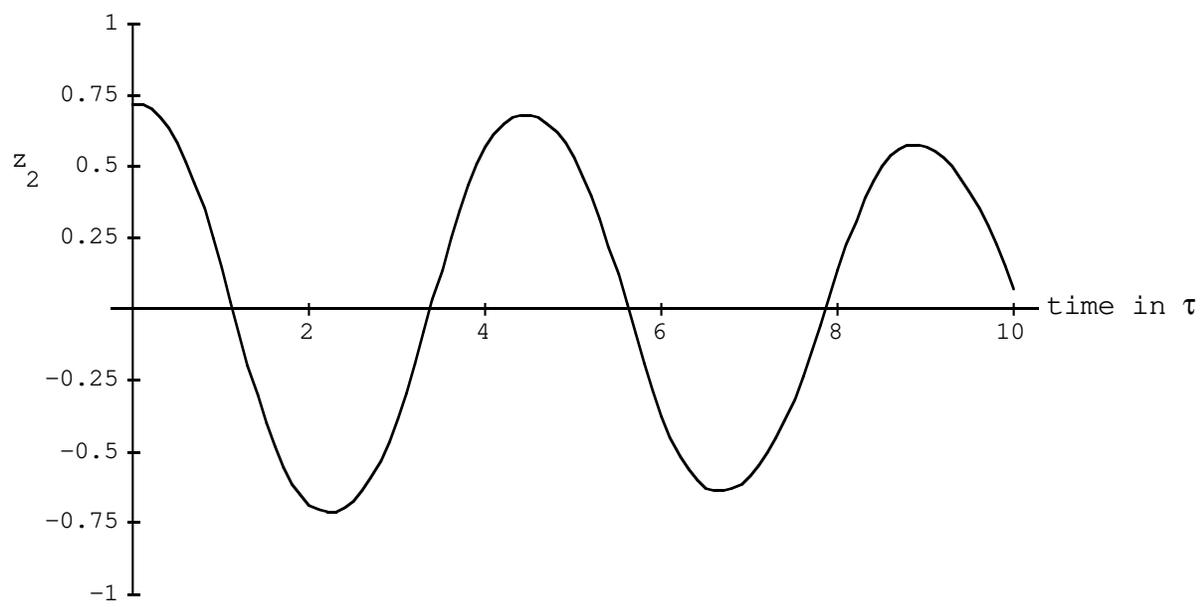
Figure 7. A comparison of $y_2$ and $z_2$ 
when $\varphi$-phase locking does not hold 
for the case 
$\stth = 0.8$, $\nubarrs = 0.9$ and  
$\gamma = \bar \gamma = 0.5$.    
(a) The component $y_2$. 
(b) The component $z_2$. 

\medskip

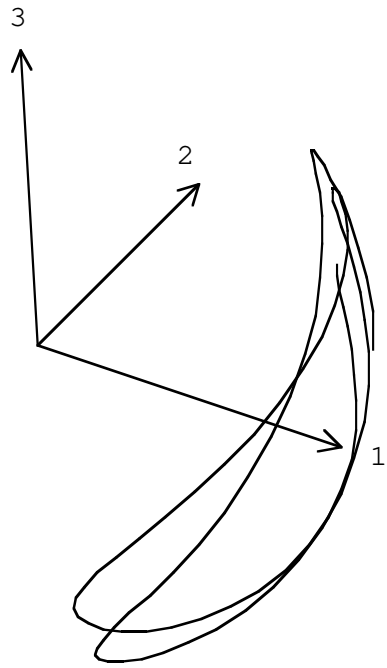
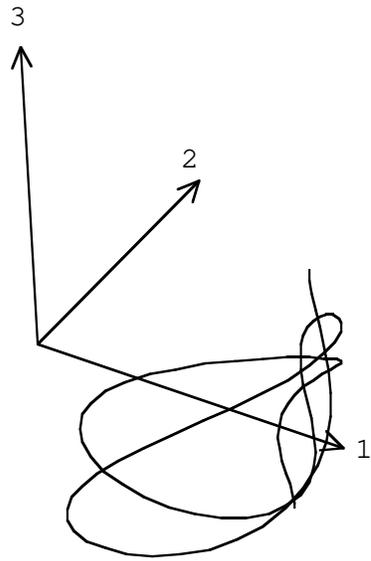
Figure 8. The orbits  
during $0 < \tau < 10$ 
for parameters as in Figure 7.      
(a) The neutrino orbit. 
(b) The antineutrino orbit.

\medskip

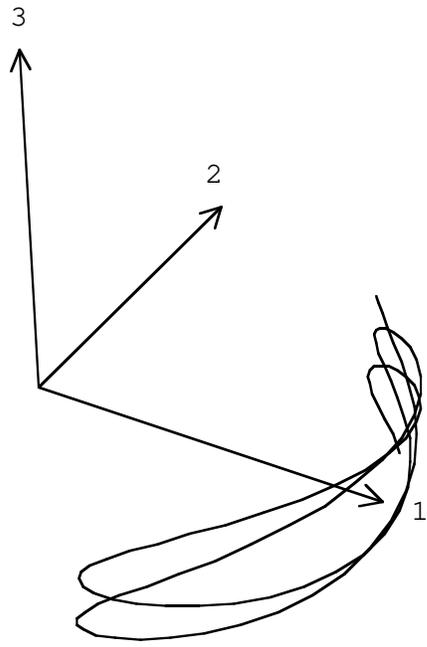
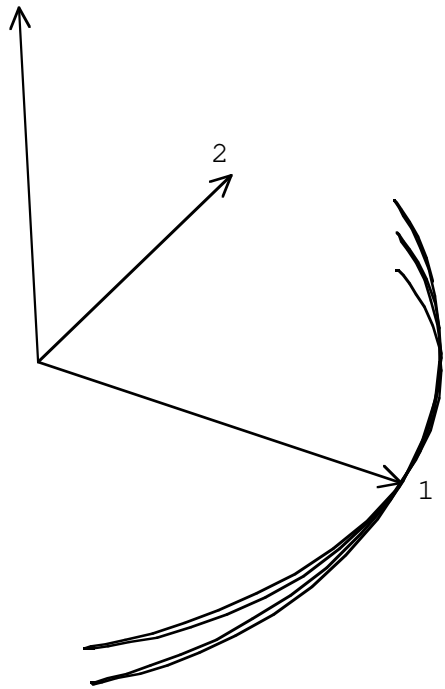
Figure 9. The orbits during 
$0 < \tau < 10$ 
for the case     
$\stth = 0.8$, $\nubarrs = 1.0$ and  
$\gamma = \bar \gamma = 0.1$  
and using the initial conditions in \eq\bref{5p1} 
with $\sin 2\alpha = 0.01$ and $\sin 2\beta = 0.0$. 
(a) The neutrino orbit. 
(b) The antineutrino orbit. 

\medskip

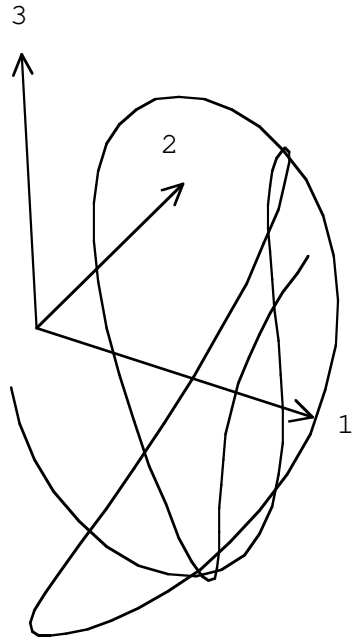
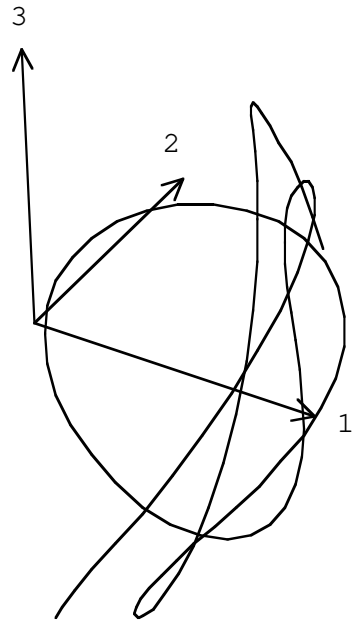
Figure 10. The orbits during 
$0 < \tau < 10$ 
for the case     
$\stth = 0.8$, $\nubarrs = 1.0$ and  
$\gamma = \bar \gamma = 0.1$  
and using the initial conditions in \eq\bref{5p1} 
with $\sin 2\alpha = \sin 2\beta = 0.1$. 
(a) The neutrino orbit. 
(b) The antineutrino orbit. 

\medskip

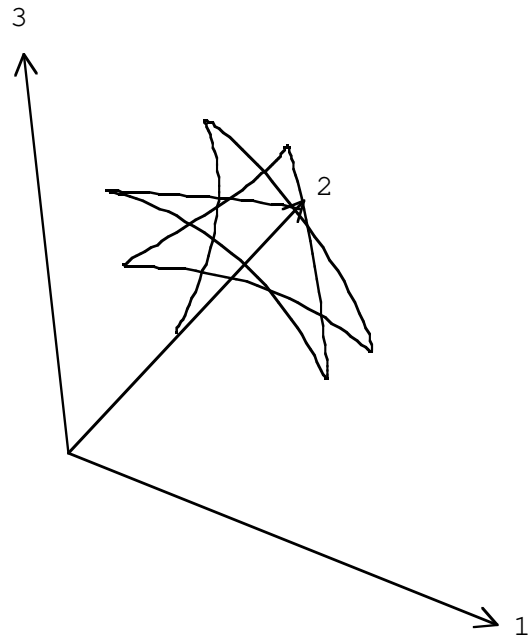
Figure 11. A neutrino orbit 
with axes in the flavor basis 
for $\Delta = - 1.0 \times 10^{-6}$ eV$^2$ and 
$\sttth = 1.0 \times 10^{-8}$ 
during a small time interval 
at approximately $0.13$ seconds after the Big Bang.  

\end{document}